\documentclass[prd,twocolumn,amsmath,amssymb,floatfix,nofootinbib]{revtex4}
\usepackage{mathrsfs,bm}
\usepackage{longtable,lscape}
\usepackage{txfonts}
\usepackage{indentfirst}
\usepackage{graphicx,color,dcolumn,booktabs}
\usepackage{multirow}
\usepackage{indentfirst}
\usepackage{multirow}
\usepackage{epsfig}
\usepackage{graphicx,color,dcolumn}
\usepackage{epstopdf}
\usepackage{amsmath}
\usepackage{multirow}
\usepackage{diagbox}
\usepackage{changes}
\usepackage{threeparttable}

\usepackage{color}
\usepackage{amssymb}
\definecolor{cover}{rgb}{0.77,0.87,0.88}
\definecolor{blueone}{rgb}{0.1,0.1,.7}
\definecolor{citec}{rgb}{0.14,0.47,0.09}
\definecolor{two}{rgb}{0.0,0.5,0.}
\definecolor{three}{rgb}{.5,.1,0.15}
\usepackage[bookmarks=true,bookmarksopen=false,plainpages=false,breaklinks=true,
   bookmarksnumbered=true,hypertexnames=false,
   filecolor=blue,urlcolor=three,menucolor=three,
   linkcolor=three,citecolor=blueone, colorlinks,
   anchorcolor=blue,runcolor=pink,frenchlinks=red
   pdfstartview=FitH,pdftitle=title,%
   pdfauthor=author]{hyperref}

\usepackage{relsize}
\usepackage{xspace}
\def\babar{\mbox{\slshape B\kern-0.1em{\smaller A}\kern-0.1em
    B\kern-0.1em{\smaller A\kern-0.2em R}}}

\begin{document}
\title{$X(2239)$ and $\eta(2225)$ as hidden-strange molecular states from $\Lambda\bar{\Lambda}$ interaction}

\author{Jun-Tao Zhu$^{1,}$\footnote{These authors have contributed equally to this work.\label{aa}}, Yi Liu$^{1,}$\footnotemark[1], Dian-Yong Chen$^{2}$, Longyu Jiang$^{3}$, Jun He$^{1,}$\footnote{Corresponding author: junhe@njnu.edu.cn}}
\affiliation{$^1$Department of  Physics and Institute of Theoretical Physics, Nanjing Normal University, Nanjing 210097, China,\\
$^2$School of Physics, Southeast University,  Nanjing 210094, China,\\
$^3$The Laboratory of Image Science and Technology, Southeast University, Nanjing 210096, China}

\date{\today}
\begin{abstract}

In this work, we propose a possible  assignment of the newly observed $X(2239)$, as well as the $\eta(2225)$,  as a molecular state from the interaction of a baryon $\Lambda$ and an antibaryon $\bar{\Lambda}$.   With the help of  effective Lagrangians, the $\Lambda\bar{\Lambda}$ interaction is described within the one-boson-exchange model with  $\eta$, $\eta'$, $\omega$, $\phi$, and $\sigma$ exchanges considered.  After inserting the potential kernel into the quasipotential Bethe-Salpeter equation, the bound states from the $\Lambda\bar{\Lambda}$ interaction can be studied by searching for the pole of  the scattering amplitude.  Two loosely bound states  with spin parities $I^G(J^{PC})=0^+(0^{-+})$ and $0^-(1^{--})$ appear near the threshold almost with the same parameter.  The $0^-(1^{--})$  state can be assigned to the $X(2239)$ observed at BESIII, which is very close to the $\Lambda\bar{\Lambda}$ threshold.  The scalar meson $\eta(2225)$ can be interpreted as a $0^+(0^{-+})$ state from the $\Lambda\bar{\Lambda}$ interaction. The annihilation effect is also discussed through a coupled-channel calculation plus a phenomenological  optical potential. It provides large widths to two  bound states produced from the $\Lambda\bar{\Lambda}$ interaction. The mass of the $1^-$ state is a little larger than the mass of the $0^-$ state after including the annihilation effect, which is consistent with our assignment of these two states as $X(2239)$ and $\eta(2225)$, respectively. The results suggest that further investigation is expected to understand the structures near the $\Lambda\bar{\Lambda}$ threshold, such as  $X(2239)$,  $\eta(2225)$,  and $X(2175)$.

\end{abstract}

\maketitle
\section{INTRODUCTION}

After the observation of $X(3872)$ at Belle, more and more $XYZ$ particles
were reported at different experimental facilities,  and attract great
interest from  theoretical side~\cite{Tanabashi:2018oca}.  Many $XYZ$
particles are suggested to be candidates of the exotic hadrons beyond the
conventional $q\bar{q}/qqq$ picture.  One of the popular interpretations of
the $XYZ$ particles  is the molecular state, which is a loosely bound state
composed of two hadrons.  The possible molecular states are widely
discussed theoretically and applied to explain the observed exotic hadrons.
The molecular states from the interaction of charmed/bottomed and
anticharmed/antibottomed mesons  are often related to the $XYZ$ particles,
such as the $Z_c(3900)$, $Z_b(4020)$, $Z_b(10610)$, and
$Z_c(10650)$~\cite{Wang:2013cya,Guo:2014iya,Aceti:2014uea,He:2015mja,He:2013nwa,Wang:2014gwa,Sun:2011uh}.
The recent observed $P_c$ states near the $\Sigma_c^{(*)}\bar{D}^{(*)}$
threshold give people more confidence in the molecular state
picture~\cite{Wu:2010jy,
Yang:2011wz,Chen:2015loa,Roca:2015dva,He:2015cea,Karliner:2015ina,Liu:2019tjn,He:2019ify,
Xiao:2019mst, Wu:2019rog,Meng:2019ilv,Wang:2019ato}.  In the light sector, the $\Lambda(1405)$ is
also proposed to be generated from the $\bar{K}N$
interaction~\cite{Oller:2000fj,Oset:1997it,Hall:2014uca,He:2015cca}.
However, the study of a molecular state composed of a baryon and an
antibaryon is scarce in the literature, and the experimental hint about
such state was also  rarely reported.  In  the charmed sector,  the
$Y(4630)$ was explained as a bound sate from the $\Lambda_c\bar{\Lambda}_c$
interaction~\cite{Lee:2011rka}. Theoretically, the interaction
between two baryons is analogous to that between two mesons. Moreover,
generally speaking, a baryon-antibaryon pair is also not difficult to be
produced in experiment. Hence, it is interesting to study the molecular
state composed of a  baryon and an antibaryon.

In fact, even before  proposition of  the quark model, the possibility to
interpret  $\pi$ meson as a $N\bar{N}$ bound state  was  discussed by Fermi
and Yang~\cite{Fermi:1959sa}. However, such attempt is incorrect based on
later studies, and was abandoned soon.  The $X(1835)$ was also connected to a $N\bar{N}$
bound state~\cite{Datta:2003iy,Yan:2004xs,Loiseau:2005cv}.  Recently, the
BESIII collaboration reported a resonance structure by analyzing the cross
section of the process $e^{+}e^{-} \rightarrow K^{+}K^{-}$  at the
center-of-mass energies ranging from $2$ to $3.08$~GeV. The structure is
denoted as $X(2239)$ which has a mass of $2239\pm7.1\pm11.3$ MeV and a
width of $139.8\pm12.3\pm20.6$ MeV~\cite{Ablikim:2018iyx}. Some
investigations were performed to interpret the
$X(2239)$~\cite{Azizi:2019ecm,Lu:2019ira}. In Ref.~\cite{Lu:2019ira}, based
on the mass estimated in a relativized quark model,  the $X(2239)$ can be
explained as a candidate of  P-wave $ss\bar{s}\bar{s}$ tetraquark state.
An important observation about the $X(2239)$ is that it is almost at the
threshold of the $\Lambda\bar{\Lambda}$ interaction after considering the
experimental uncertainty of the mass. If we recall that the $X(2239)$ has
spin parity  $J^{P}=1^{-}$ and  was observed in the hidden-strange $K^+K^-$
channel, it is a good candidate for a hidden-strange molecular state
composed of a baryon $\Lambda$ and an antibaryon $\bar{\Lambda}$.

Before the observation of $X(2239)$,  another state with the same quantum number, the $Y(2175)$, also named as $\phi(2170)$ in the literature, was
observed by the Babar Collaboration in the initial-state-radiation process
$e^{+}e^{-}\rightarrow\gamma_{ISR}\phi(1020)f_{0}(980)$ with a mass of
about $2175$ MeV~\cite{Aubert:2006bu}. Since the $Y(2175)$ was observed, it
has been investigated in many theoretical pictures, which include $qqg$
hybrid~\cite{Ding:2006ya,Ding:2007pc}, $ss\bar{s}\bar{s}$ tetraquark
state~\cite{Wang:2006ri,Chen:2008ej,Drenska:2008gr}, excited
$1^{--}s\bar{s}$ state~\cite{Shen:2009zze}, resonance state of $\varphi
K\bar{K} $~\cite{MartinezTorres:2008gy,GomezAvila:2007ru}, and some other
interesting speculations~\cite{Zhu:2007wz,AlvarezRuso:2009xn,Yuan:2008br}.
It is also possible that the $X(2239)$ and $X(2175)$ are the same
state~\cite{Tanabashi:2018oca,Chen:2020xho}.  However,  a  $1^{--}$ state with a mass of
$2135\pm8\pm9$~MeV and a width of $104\pm24\pm12$~MeV was also observed in
the $\phi f_0(980)$ channel at BESIII~\cite{Ablikim:2017auj}.  It is more
appropriate to take these two states as two separate states if we accept
the large mass gap of these two states as observed experimentally.
Besides, a state with a mass of about $2220$ MeV was observed by the DM2
Collaboration and  confirmed at MARK-III  in the radiative decays
$J/\psi\rightarrow\gamma\phi\phi$~\cite{Bisello:1986pt,Bai:1990hk}. Later,
the BES  and BESIII Collaborations also confirmed the  existence of the
$\eta(2225)$~\cite{Ablikim:2008ac,Ablikim:2016hlu}. There exist also a few
theoretical interpretations of the $\eta(2225)$, such as a $4^{1}S_{0}$
$s\bar{s}$ state~\cite{Li:2008we,Wang:2017iai}.

As indicated in Ref.~\cite{Zhao:2013ffn}, which was done before the
observation of  $X(2239)$ at BESIII,  the $Y(2175)$  and $\eta(2225)$ can
be interpreted as  $\Lambda\bar{\Lambda}(^{3}S_{1})$ and
$\Lambda\bar{\Lambda}(^{1}S_{0})$ molecular states, respectively, which is
also the first attempt to discuss the possible molecular state from the
$\Lambda\bar{\Lambda}$ interaction. However, it should be noticed that the
$\Lambda\bar{\Lambda}$ threshold is about 2231.3~MeV,  while the mass of
$Y(2175)$ is about $60$ MeV lower than the $\Lambda\bar{\Lambda}$
threshold, which is too deep to be a molecular state. Moreover, a recent
measurement at BESIII indicates that the mass of $Y(2175)$  is about 2135
MeV~\cite{Ablikim:2017auj}, which is about one hundred MeV below the
$\Lambda \bar{\Lambda}$ threshold.  As for the  newly observed  $X(2239)$,
its mass seems closer to the  $\Lambda\bar{\Lambda}$ threshold. Hence, it
is interesting to study the possibility of assignment of the $X(2239)$,
rather than the $Y(2175)$, as candidate of $\Lambda\bar{\Lambda}(1^{-})$
molecular state. There also exist theoretically study about the molecular
state from the $\Lambda{\Lambda}$
interaction~\cite{Sasaki:2019qnh,Nagels:2015dia,Haidenbauer:2015zqb}. It is
found that in a lattice calculation the $\Lambda{\Lambda}$ interaction is
attractive, but too weak to form a molecular state~\cite{Sasaki:2019qnh}.

Recalling the results in Ref.~\cite{Zhao:2013ffn}, one can find that the
mass gap between the $Y(2175)$ and $\eta(2235)$ was  reproduced from a
calculation with S-wave $\Lambda\bar{\Lambda}$ interaction in the
one-boson-exchange model  by solving the non-relativistic Schr\"odinger
equation. The $1^-$ state has a larger binding energy than $0^-$ state. And the D-wave interaction only involve in the $1^-$ state,  which suggests that inclusion of the relativistic effect and the S-D mixing may change the mass gap between two states.  In the quasipotential Bethe-Salpeter (qBSE) approach, such effects can be included naturally.  It is interesting to make a calculation in such approach to see  the variation of the mass gap. It provides a possibility to obtain two bound states both close to the threshold, which is more consistent with the molecular state as a loosely bound state of two hadrons.  Besides, for the $N\bar{N}$ interaction, the annihilation effect was found important in the
literature~\cite{Phillips:1967zza,Klempt:2005pp,Richard:2019dic,Amsler:1997up}, which may affect  the $\Lambda\bar{\Lambda}$ interaction  also. The theoretical values of the mass will deviate
from the one within the one-boson-exchange model, and the bound state will
acquire a width after the annihilation effect is included. 

In the current work, we  adopt qBSE approach to study the $\Lambda\bar{\Lambda}$ interaction.  With the help of the effective Lagrangians, the one-boson-exchange model with pseudoscalar,  scalar, and vector exchanges is applied to construct the interaction. The
annihilation effect will be introduced by  coupled-channel effect plus
an imaginary optical potential. 
By inserting  the potential into  the qBSE, the molecular states
with quantum numbers  $I^G(J^{PC})=0^+(0^{-+})$ and $0^-(1^{--})$ will be investigated.

The paper is organized as follows. After the introduction, we present
relevant Lagrangians to construct the meson exchange potential. The qBSE
approach is also briefly introduced in Section \ref{Sec: Formalism}.  The
numerical results of bound states produced from the $\Lambda\bar{\Lambda}$
interaction within one-boson-exchange model are presented in Subsection \ref{OBE}. We discuss
the annihilation effect on the $\Lambda\bar{\Lambda}$ interaction, and the
results with such effect are given in  Subsection~\ref{ann}.  The paper ends
with discussion and summary.

\section{Theoretical frame}\label{Sec: Formalism}

%{\color{red}Lagrangians including coupling constant }\\
First, we describe  the $\Lambda\bar{\Lambda}$ interaction within the one-boson-exchange model. As in Ref.~\cite{Zhao:2013ffn},  to construct the potential,  the Lagrangians for the couplings between the $\Lambda$ baryon and exchanged mesons  can be written as,
%ae??ae¡ã?šŠ??
\begin{eqnarray}
\mathcal{L}_{\eta\Lambda\Lambda}&=& -ig_{\eta\Lambda\Lambda}\bar{\psi}_{\Lambda}\gamma_{5}\psi_{\Lambda}\eta,\label{L1}\\
\mathcal{L}_{\eta'\Lambda\Lambda}&=& -ig_{\eta'\Lambda\Lambda}\bar{\psi}_{\Lambda}\gamma_{5}\psi_{\Lambda}\eta',\\
\mathcal{L}_{\sigma\Lambda\Lambda}&=& g_{\sigma\Lambda\Lambda}\bar{\psi}_{\Lambda}\psi_{\Lambda}\sigma,\\
\mathcal{L}_{\omega\Lambda\Lambda}&=& -g_{\omega\Lambda\Lambda}\bar{\psi}_{\Lambda}\gamma_{\mu}\omega^{\mu}\psi_{\Lambda},\\
%+\frac{f_{\omega\Lambda\Lambda}}{2m_{\Lambda}}\bar{\psi}_{\Lambda}\sigma_{\mu\nu}\psi_{\Lambda}\partial^{\mu}\omega_{\nu}\\
\mathcal{L}_{\phi\Lambda\Lambda}&=& -g_{\phi\Lambda\Lambda}\bar{\psi}_{\Lambda}\gamma_{\mu}\phi^{\mu}\psi_{\Lambda},\label{L5}
%+\frac{f_{\phi\Lambda\Lambda}}{2m_{\Lambda}}\bar{\psi}_{\Lambda}\sigma_{\mu\nu}\psi_{\Lambda}\partial^{\mu}\phi_{\nu}
\end{eqnarray}
where the $\psi_{\Lambda}$, $\eta$, $\eta'$, $\sigma$, $\omega$, and $\phi$ are the fields of $\Lambda$ baryon, $\eta$, $\eta'$, $\sigma$, $\omega$, and $\phi$ mesons. The coupling constant $g_{\alpha\Lambda\Lambda}$ can be derived by the SU(3) symmetry and considering the mixings between octet and singlet states~\cite{Zhao:2013ffn},  and the masses of  exchanged mesons $m_e$ are cited from the Review of Particle Physics (PDG)~\cite{Tanabashi:2018oca}, the explicit values are listed below,
%šš??šŠ??ae?¡¥šš????????ae?¡ã
\begin{eqnarray}
g^{2}_{\eta\Lambda\Lambda}/4\pi&=& 4.473 ,  \quad
 m_{\eta}=548.8~{\rm MeV},\notag\\
g^{2}_{\eta'\Lambda\Lambda}/4\pi&=& 9.831 , \quad
 m_{\eta'}=957.7~{\rm MeV},\notag\\
g^{2}_{\sigma\Lambda\Lambda}/4\pi&=& 3.459, \quad
 m_{\sigma}=500.0~{\rm MeV},\notag\\
g^{2}_{\omega\Lambda\Lambda}/4\pi&=& 8.889 , \quad
 m_{\omega}=782.6~{\rm MeV},\notag\\
g^{2}_{\phi\Lambda\Lambda}/4\pi&=& 2.222,  \quad
  m_{\phi}=1019.5~{\rm MeV}\nonumber.
\end{eqnarray}
The mass of $\sigma$ meson has a large uncertainty $400-550$ MeV~\cite{Tanabashi:2018oca}. Here we choose a value of 500 MeV, and the uncertainty will be discussed latter.

In the current work, we consider the $\Lambda\bar{\Lambda}$ interaction instead of the $\Lambda\Lambda$ interaction. Hence, the couplings between the light mesons and the antibaryon $\bar{\Lambda}$ are also required.  As in the nucleon-antinucleon interaction, we  adopt the well-known G-parity rule to write the  $\Lambda\bar{\Lambda}$ interaction from the $\Lambda\Lambda$ interaction. By inserting the $G^{-1}G$ operator into the potential, the G-parity rule can be obtained easily as~\cite{Phillips:1967zza,Klempt:2002ap},
\begin{eqnarray}
V&=&\sum_{i}{\zeta_{i}V_{i\Lambda\Lambda}}.
\end{eqnarray}
The G parity of the exchanged meson is left as a $\zeta_{i}$ factor for $i$ meson. Since $\omega$ and $\phi$ mesons carry odd $G$ parity, $\zeta_{\omega}$ and $\zeta_{\phi}$ should equal $-1$, and others still equal $1$. Finally, we reach a relation as,
\begin{eqnarray}
V_{\Lambda\bar{\Lambda}}&=& V_{\eta\Lambda\Lambda}+V_{\eta'\Lambda\Lambda}+V_{\sigma\Lambda\Lambda}-V_{\omega\Lambda\Lambda}-V_{\phi\Lambda\Lambda}.\label{V0}
\end{eqnarray}

Now, we only need  the potential of the $\Lambda{\Lambda}$ interaction.  With the Lagrangians and the coupling constants given above, we can write the relevant meson exchange potentials  with the standard Feynman rule as,
\begin{eqnarray}
iV_{\mathbb{P}\Lambda\Lambda} &=& -g^{2}_{\mathbb{P}\Lambda\Lambda}\bar{u}_{\Lambda}\gamma_{5}u_{\Lambda}
  \frac{1}{q^{2}-m^{2}_{\mathbb{P}}}f_{i}(q^{2})\bar{u}_{\Lambda}\gamma_{5}u_{\Lambda},\notag\\
iV_{\mathbb{V}\Lambda\Lambda} &=&g^{2}_{\mathbb{V}\Lambda\Lambda}\bar{u}_{\Lambda}\gamma_{\mu}u_{\Lambda}
  \frac{-g^{\mu\nu}+q^{\mu}q^{\nu}/m^{2}_{\mathbb{V}}}{q^{2}-m^{2}_{\mathbb{V}}}f_{i}(q^{2})\bar{u}_{\Lambda}\gamma_{\mu}u_{\Lambda},\notag\\
iV_{\sigma\Lambda\Lambda} &=& g^{2}_{\sigma\Lambda\Lambda}\bar{u}_{\Lambda}u_{\Lambda}
  \frac{1}{q^{2}-m^{2}_{\sigma}}f_{i}(q^{2})\bar{u}_{\Lambda}u_{\Lambda},
\end{eqnarray}
where the $u_\Lambda$ is the spinor of the $\Lambda$ baryon. The $q$, $m_\mathbb{P}$, $m_\mathbb{V}$, and $m_\sigma$ are the exchanged momentum, and the masses of exchanged pseudoscalar $\mathbb{P}$ ($\eta$ and $\eta'$), vector $\mathbb{V}$ ($\omega$ and $\phi$) and scalar $\sigma$ mesons.

Usually, a form factor should be introduced at the vertices because  the exchanged mesons are not   point particles and have internal structure.  Such form factors are also used to ensure the convergence of the integral (the qBSE is an integral equation and will be given later).  There exist many types of the form factors in the literature. Due to  absence of  experimental data of the $\Lambda\Lambda$ interaction, we can not determine which type is more realistic.   In the current work, we adopt three types of  form factors as in Ref.~\cite{He:2019rva},
%?oe¡é????? ?-?
 \begin{eqnarray}
   f_{1}(q^{2}) &=& \frac{\Lambda^{2}_{e}-m^{2}_{e}}{\Lambda^{2}_{e}-q^{2}}, \label{f1}\\
   f_{2}(q^{2}) &=& \frac{\Lambda^{4}_{e}}{(m^{2}_{e}-q^{2})^2+\Lambda^{4}_{e}} ,\label{f2}\\
   f_{3}(q^{2}) &=& e^{-(m^{2}_{e}-q^{2})^2/\Lambda^{4}_{e}}\label{f3}.
 \end{eqnarray}
 We  parameterize the cutoff in a form of $\Lambda_{e}=m_{e}+\alpha_{e}~0.22$~GeV with $m_e$ being  the mass of  exchanged meson~\cite{He:2019rva,He:2019ify,Cheng:2004ru,Chen:2012nva,Guo:2016iej,Yu:2017zst}.   Such parameterization of  cutoff can introduce the effect of the mass of exchanged meson, which is more reasonable than the adoption of the same cutoff  for different mesons.  The $\alpha_{e}$ is taken as a free parameter which is close to 1.  Considering the explicit forms of  form factors, the $\alpha$ for $f_1$ should be larger than 0 to avoid an unphysical suppression near $\Lambda_e=m_e$.  For the other two types of  form factors, a value about zero can be chosen.
The above form factors satisfy the requirement as $f(m^2)=1$.  And the radius of  $\Lambda$ baryon can be estimated with a relation $r^2=6/f(0)~df(q^2)/dq^2|_{q^2\to0}$, which leads to a reasonable value about 0.5 fm for three choices of form factors.  Hence, the three types of form factors satisfy the basic requirements, and we will further check  whether our conclusion is sensitive to  different choices.

Different from Ref.~\cite{Zhao:2013ffn}, we will adopt the qBSE to explore  possible bound states from the $\Lambda\bar{\Lambda}$ interaction.  The potential kernel obtained above will be inserted  into  the Bethe-Salpeter equation to obtain the scattering amplitude, the poles of which correspond to  bound states. The Bethe-Salpeter equation is a 4-dimensional integral equation in the Minkowski space.  Considering the complexity and  difficulty of directly solving such integral equation, we adopt a quasipotential approximation approach to reduce the 4-dimensional Bethe-Saltpeter equation into a 3-dimensional integral equation~\cite{Gross:2010qm,He:2012zd,He:2011ed}. Then, using the partial-wave decomposition, the 3-dimensional equation is further reduced into a 1-dimensional equation with fixed spin parity $J^P$ as~\cite{He:2015mja,He:2017aps},
\begin{eqnarray}
i{\cal M}^{J^P}_{\lambda'\lambda}({\rm p}',{\rm p})
&=&i{\cal V}^{J^P}_{\lambda',\lambda}({\rm p}',{\rm
p})+\sum_{\lambda''}\int\frac{{\rm
p}''^2d{\rm p}''}{(2\pi)^3}\nonumber\\
&\cdot&
i{\cal V}^{J^P}_{\lambda'\lambda''}({\rm p}',{\rm p}'')
G_0({\rm p}'')i{\cal M}^{J^P}_{\lambda''\lambda}({\rm p}'',{\rm
p}),\quad\quad \label{Eq: BS_PWA}
\end{eqnarray}
where the sum extends only over nonnegative helicity $\lambda''$. In the current case, we will consider spin parities $J^P=0^-$ and $1^-$,  which can couple to baryons $\Lambda$ and $\bar{\Lambda}$ in S wave, and are called S-wave states in the non-relativistic calculation~\cite{Zhao:2013ffn}. Since we make a decomposition on spin parity $J^P$ directly, contributions from all possible orbital angular momenta $L$  are included naturally.  Hence,  in the qBSE approach, no special treatment is needed to include the D-wave contribution.

The reduced propagator with the spectator approximation can be written down
in the center-of-mass frame with $P=(W,{\bm 0})$ as,
\begin{align}
	G_0&=\frac{\delta^+(p''^{~2}_2-m_2^{2})}{p''^{~2}_1-m_1^{2}}
	\nonumber\\&=\frac{\delta(p''^{0}_2-E_2({\rm p}''))}{2E_2({\rm p''})[(W-E_2({\rm
p}''))^2-E_1^{2}({\rm p}'')]},
\end{align}
where the $\delta^+(p''^{~2}_2-m_2^{2})$ is the dirac delta function but
with only $p''^{0}_2=+E_2({\rm p}'')$. Here, as required by the spectator approximation, we put one of the
particles, 2 here, on shell, which satisfies  $p''^0_2=E_{2}({\rm
p}'')=\sqrt{ m_{2}^{~2}+\rm p''^2}$. In above equations, a definition ${\rm p}=|{\bm p}|$
is adopted.

The partial-wave potential is defined with the potential of the $\Lambda\bar{\Lambda}$ interaction obtained in the above as,
\begin{eqnarray}
{\cal V}_{\lambda'\lambda}^{J^P}({\rm p}',{\rm p})
&=&2\pi\int d\cos\theta
~[d^{J}_{\lambda\lambda'}(\theta)
{\cal V}_{\lambda'\lambda}({\bm p}',{\bm p})\nonumber\\
&+&\eta d^{J}_{-\lambda\lambda'}(\theta)
{\cal V}_{\lambda'-\lambda}({\bm p}',{\bm p})],
\end{eqnarray}
where $\eta=PP_1P_2(-1)^{J-J_1-J_2}$ with $P$ and $J$ being parity and spin for the system, $\Lambda$, or $\bar{\Lambda}$ baryon. The initial and final relative momenta are chosen as ${\bm p}=(0,0,{\rm p})$  and ${\bm p}'=({\rm p}'\sin\theta,0,{\rm p}'\cos\theta)$. The $d^J_{\lambda\lambda'}(\theta)$ is the Wigner d-matrix.
Since particle 1 is off-shell in the qBSE approach, a form factor should be also introduced to reflect its internal structure.  Here, we  adopt an exponential regularization by introducing a form factor into the propagator as
$G_{0}(p)\rightarrow G_{0}(p)[e^{-(k^{2}_{1}-m^{2}_{1})^{2}/\Lambda^{4}_{r}}]^{2}$
where the $k_{1}$ and  $m_{1}$ are the momentum and the mass of the off-shell particle. With such regularization, the convergence of the integral equation is guaranteed even without the form factor for  exchanged meson.  The cutoff $\Lambda_{r}$ is parameterized as in the $\Lambda_{e}$ case, that is,  $\Lambda_{r}=m_{e}+\alpha_{r}~0.22$~GeV with $m_{e}$ being the mass of  exchanged meson and $\alpha_{r}$ serving the same function as the parameter $\alpha_{e}$.

With the Gauss discretization of momentum,  the 1-dimensional integral equation in Eq.~(\ref{Eq: BS_PWA}) is transformed into a matrix equation as $M=V+VG_0M$~\cite{He:2015cca}. The molecular states correspond to the poles of scattering  amplitude $ M$ in complex energy plane at $|1-V(z)G(z)|=0$ with $z=W+i\Gamma/2$ being  system energy $W$ at real axis~\cite{He:2015mja}. 

\section{Numerical results}\label{Sec: results}

We first consider the case without the annihilation effect included.  Since only one channel is considered in
this work, the bound state pole is  located at real axis.  The parameters
in the qBSE approach are the cutoffs $\Lambda_{e,r}$ which have been
parameterized into $\alpha_{e,r}$.  In the calculation, we choose
$\alpha_e$ equivalent  to $\alpha_r$, and rename them as a parameter
$\alpha$,  for simplification.  We consider two spin parities $0^-$ and
$1^-$, which can be obtained from S-wave coupling.

\subsection{Bound states from  $\Lambda\bar{\Lambda}$ interaction}
\label{OBE}

\begin{figure}[h!]
  \centering
  \includegraphics[bb=20 0 700 730,scale=0.35,clip]{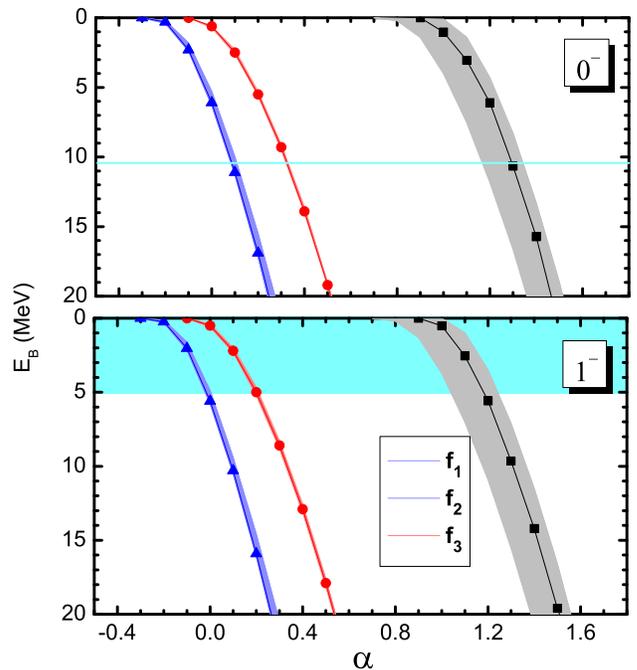}
  \caption{The binding energy $E_{B}$ with the variation of the $\alpha$.
  The upper and lower panels are for bound states with spin parities $0^-$
and $1^-$, respectively.  The black square, red circle, and blue triangle
are  for the results with different types of form factors $f_{i}(q^2)$ with
$i=1,2,3$  in Eqs.~(\ref{f1}-\ref{f3}), respectively. The bands are for the uncertainties from the uncertainties of mass of the $\sigma$ meson, $400\sim550$ MeV~\cite{Tanabashi:2018oca}. The cyan line in the
upper panel is the suggested value of mass of  $\eta(2225)$ in the
PDG~\cite{Tanabashi:2018oca}.  The cyan band in the lower panel is for the experimental mass of $X(2239)$  with uncertainties, and here only the part below the threshold is presented~\cite{Ablikim:2018iyx}. More explanations are given in the text.  }\label{total}
  \end{figure}

In Fig.\ref{total}, the binding energy $E_B= m_{th}-W$  with $m_{th}$ and $W$ being the threshold and the position of the pole obtained with different types of form factors are presented.
The bound state from the $\Lambda\bar{\Lambda}$ interaction with  quantum
numbers $I^G(J^{PC})=0^+(0^{-+})$ is presented in the upper panel of  Fig.\ref{total}. The bound state can be produced from the interaction with reasonable $\alpha$.  For the monopole type of the form factors $f_1(a^2)$, the bound state appears at an $\alpha$ of about 1, which is a standard value of the $\alpha$.  In  Fig.\ref{total}, the suggested value of the mass of $\eta(2225)$ in the PDG~\cite{Tanabashi:2018oca} is also shown as a cyan line, which can be reproduced at an $\alpha$ of about 1.2.  Since the uncertainty of the mass of $\eta(2225)$ is about 10 MeV, which just fills the region we considered, we do not show the uncertainty in the figure. The uncertainty corresponds to a range of $\alpha$ from 0.8 to 1.5.    For other two types of form factors, the bound state is produced at $\alpha$ of about zero, which corresponds to a standard cutoff about 1 GeV.
The shapes of three curves for different form factors are analogous to each other. Considering that  $\alpha$ is a free parameter in a reasonable range,  one can say that the different choices of  form factors do not affect on the conclusion. Hence, comparing the theoretical results with  experiment, the $0^-$ state from the $\Lambda\bar{\Lambda}$ interaction can be related to the $\eta(2225)$.

Now we turn to the $0^-(1^{--})$ case, which is shown in the lower panel of
Fig.\ref{total}.  Contrary to the results in~\cite{Zhao:2013ffn},  the
binding energies of $1^-$ state are similar to these of $0^-$ state
with the same parameter.  For the monopole form factor $f_1(q^2)$, the
bound state appears at an $\alpha$ of about 0.9, increases with the
increase of $\alpha$, and reaches a binding energy about 20 MeV at an
$\alpha$ of about 1.5.  For other two types of form factors, the bound
state is produced   at $\alpha$ of about zero.  The experimental mass
of the $X(2239)$ reported by BESIII Collaboration is
$2239\pm7.1\pm11.3$ MeV. The central value is lightly higher than the
$\Lambda\bar{\Lambda}$ threshold.  After the uncertainty considered,
the $X(2239)$ is just on the threshold. In  Fig.\ref{total}, we also
present  the uncertainty of the  mass of  $X(2239)$ below the
$\Lambda\bar{\Lambda}$ threshold as a cyan band.  The experimental
uncertainty of the $X(2239)$ corresponds to $\alpha$ from 0.8 to 1.2.

From  above results, one can find that the mass gap between
$0^-$ and $1^-$ states in our model is quite small. Two states appear almost at the same cutoff, and the mass gap at a certain $\alpha$ is only several MeV in the region considered in Fig.~\ref{total}. It is quite different from the
results in Ref.~\cite{Zhao:2013ffn}. In that work, with reasonable
cutoffs, they obtained a loosely bound state of $0^-$ with binding
energy about $7\sim13$~MeV while the $1^-$ state has larger binding
energy about $50\sim82$~MeV. The mass gap is about 50 MeV, which is much larger the one in the current work. Because the Lagrangians and  coupling
constants adopted in two works are the same, the difference should be from
the different treatments, such as different solution method, the relativistic effect, and S-D mixing.  Besides, we also present the bands from the uncertainties of the mass of $\sigma$ meson. The results are found not sensitive to this uncertainty.

\begin{figure}
  \centering
    \includegraphics[bb=40 0 750 1160,scale=0.34,clip]{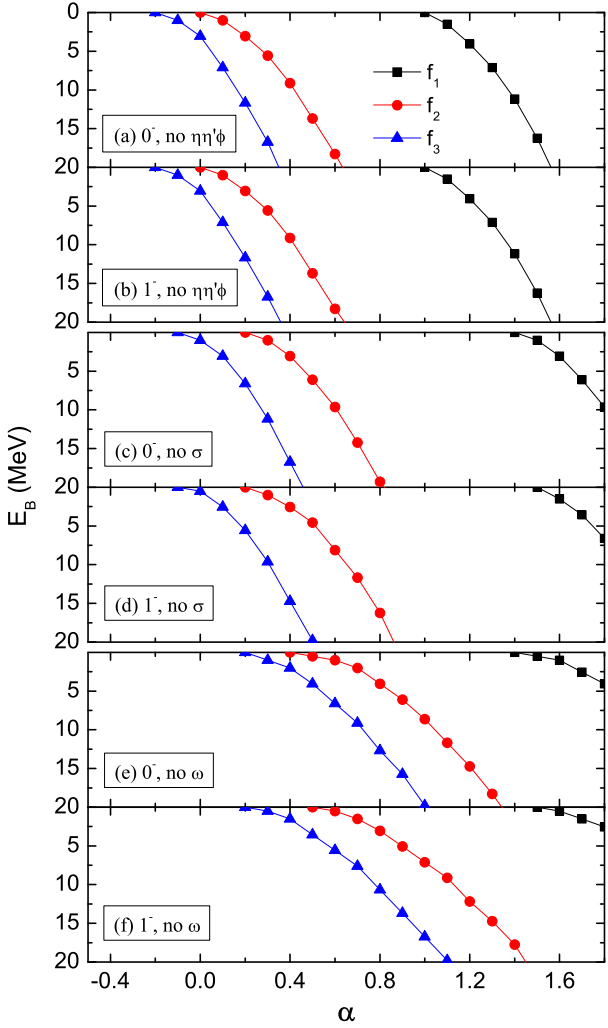}
  \caption{The binding energy $E_{B}$ without the $\eta,\eta'$ and $\phi$ exchange  (a and b),  without $\sigma$ exchange (c and d) and without $\omega$ exchange (e and f). Other
conventions are the same as in Fig. \ref{total}.}\label{role}
 \end{figure}

In our model, five exchanges including $\eta$, $\eta'$, $\phi$,
$\omega$, and $\sigma$ exchanges, are considered to construct the
interaction potential.  Usually, these exchanges play different roles
in producing  bound states. In the qBSE approach, the potential can
not be shown as a function of the range $r$ as in the non-relativistic
calculation~\cite{Zhao:2013ffn}. We check their roles by turning on
and off one or more exchanges and vary the parameter $\alpha$ from -1
to 3 to search for  bound state.  It is found that if we only keep one
of five exchanges,  no bound state can be produced from  $\eta$,
$\eta'$, or $\phi$ exchange while the interaction with only $\omega$
or only  $\sigma$ exchange is  still strong enough to produce a bound
state with a larger $\alpha$.  Such result suggests that the $\omega$
and $\sigma$ exchanges play the most important role in producing the
bound states.  In the following we give more explicit results in
Fig.~\ref{role}  to show the role of exchanges.

We present  the results after turning off $\eta$, $\eta'$ and $\phi$
exchanges and only keeping $\omega$ and $\sigma$ exchanges in panels (a)
and (b) of Fig.~\ref{role}.  As shown in the figure, the bound states with
$0^-$ and $1^-$ can be produced from the $\omega$ and $\sigma$ exchanges
with a light increase of the parameter $\alpha$ for all three types of form
factors.  We also check the case after removing both $\omega$ and $\sigma$
exchanges but keeping $\eta$, $\eta'$ and $\phi$ exchanges. No bound state
can be found with a reasonable parameter.  Such result suggests that the
$\omega$ and $\sigma$ exchanges are essential to produce the bound state
with $0^-$ and $1^-$.  In panels (c) and (d), the results after removing
$\sigma$ exchange are presented. Larger values of $\alpha$ are required for
the three types of form factors  than in the previous case.   The largest
effect comes from the $\omega$ exchange  as shown in panels (e) and (f).
To reproduce a binding energy in the full model, the parameter $\alpha$
should be increased by 0.5 or more.

\subsection{Annihilation effect from intermediated mesons}\label{ann}

In the above calculation, we do not consider the effect of the annihilation
of  baryon $\Lambda$ and antibaryon $\bar{\Lambda}$.  The annihilation effect was found
important in the study of the  $N\bar{N}$
interaction~\cite{Klempt:2005pp,Richard:2019dic,Amsler:1997up}.  Such
contribution is often considered as the multipion intermediation in $s$
channel, which is usually replaced by annihilation into two mesons,  plus
an optical
potential~\cite{Mull:1991rs,Hippchen:1991rr,Dover:1980pd,Pignone:1994uj}.
The annihilation effect  induces an imaginary potential, which leads to a width
and variation of the mass  of the bound
state~\cite{ElBennich:2008vk,Carbonell:1989cs}. In the
$\Lambda\bar{\Lambda}$ interaction,  such annihilation can occur also,
which will effect the experimental
observables~\cite{Haidenbauer:1991kt,Haidenbauer:1992wp}. 

In the literature, the two-meson intermediation part of the annihilation
effect was included by introducing box diagram or coupled-channel
effect~\cite{Shapiro:1988bv,Liu:1990rd,Mull:1994gz,ElBennich:2008vk,Carbonell:1989cs}.
In the current work we will adopt the latter treatment, i.e., a
coupled-channel calculation in our qBSE approach which was developed in
Ref.~\cite{He:2019rva}. Explicitly, we follow the method in
Ref.~\cite{Mull:1994gz}, which is
successful applied to $N\bar{N}$ interaction , and is more consistent with our qBSE
approach. 

In Ref.~\cite{Mull:1994gz}, the annihilation effect was introduced by the
two-meson intermediation  and an imaginary  phenomenological optical
potential.  For the former,  we still adopt two-meson intermediation picture here as
in the $N\bar{N}$ case. As in the above calculation, only the pseudoscalar
mesons $\mathbb{P}$ ($\pi$, $\eta$ and $\eta'$), vector mesons $\mathbb{V}$
($\omega$ and $\phi$), and scalar meson $\sigma$ will be considered to
avoid more uncertainties  from more Lagrangians and coupling constants
involved.  In the current work, we focus on states with quantum numbers
$I^G(J^{PC})=0^+(0^{-+})$ and $0^-(1^{--})$. For the former state, the
possible intermediated two-meson channels include $\mathbb{V}\mathbb{V}$
and $\mathbb{P}\sigma$, which leads to an eight-channel calculation. For
the latter state, the $\mathbb{P}\mathbb{V}$ and $\mathbb{V}\sigma$
channels involve in the calculation, which includes twelve channels. Beside, the $K\bar{K}$ channel will be considered for $0^-(1^{--})$ state, which is forbidden for $0^+(0^{-+})$ state.   We introduce  the $\Lambda\bar{\Lambda}-m_1m_2$ interaction,
where $m_1m_2$ are two of the mesons considered.  Following the treatment
in the $N\bar{N}$ case~\cite{Pignone:1994uj}, all interactions between two
mesons and couplings between different  meson channels are ignored in the
calculation. 

As in Ref.~\cite{Mull:1994gz}, here, we take a two-channel
interaction to give a simple explanation about the  relation of the
standard coupled-channel approach to the well-known forms in the study of
the  the annihilation $N\bar{N}$ interaction from box diagram in
Ref.~\cite{ElBennich:2008vk}.
The coupled-channel Bethe-Salpeter equation in matrix form is written as
\begin{align}
&\left(\begin{array}{cc}
M^{BB}&M^{Bm}\\
M^{mB}&M^{mm}
\end{array}
\right)=
\left(\begin{array}{cc}
V_{el}&V^{Bm}\\
V^{mB}&0
\end{array}
\right)\nonumber\\&+
\left(\begin{array}{cc}
V_{el}&V^{Bm}\\
V^{mB}&0
\end{array}
\right)
\left(\begin{array}{cc}
G^{BB}&0\\
0&G^{mm}
\end{array}
\right)\left(\begin{array}{cc}
M^{BB}&M^{Bm}\\
M^{mB}&M^{mm}
\end{array}
\right),
\end{align}
where $B$ and $m$ mean $\Lambda\bar{\Lambda}$ and $m_1m_2$ channels,
respectively, and $V_{el}$ is the potential given in Eq.~(\ref{V0}). Here,
we choose $V^{mm}=0$, that is, the interaction  between two mesons are not
considered. Then we can obtain following equations
\begin{align}
M^{BB}&=V_{el}+V_{el}G^{BB}M^{BB}+V^{Bm}G^{mm}M^{mB},\label{aa}\\
M^{mB}&=V^{mB}+V^{mB}G^{BB}M^{BB}.\label{bb}
\end{align}
By inserting Eq.~(\ref{bb}) into Eq.(\ref{aa}), we obtian
\begin{align}
M^{BB}&=V^{BB}+V^{BB}G^{BB}M^{BB},\\
M^{mB}&=V^{mB}+V^{mB}G^{BB}M^{BB},
\end{align}
where we  define  $V^{BB}=V_{el}+V^{Bm}G^{mm}V^{mB}$ as in Ref.~\cite{Mull:1994gz}. Here the second term is the annihilation term from a box diagram.
Hence, we need the transition potential which can be obtained from the Lagrangians in Eqs.~(\ref{L1}-\ref{L5}) as
\begin{eqnarray}
{V}^{mB} &=&\zeta g_{m_1\Lambda\Lambda}g_{m_2\Lambda\Lambda}\bar{u}_{\Lambda}\Gamma_1
  \frac{ q\!\!\!/+m_\Lambda}{q^{2}-m^{2}_{\Lambda}}f_i(q^2)\Gamma_2 v_{\bar\Lambda},
\end{eqnarray}
where $g_{m_{1,2}\Lambda\Lambda}$ is coupling constant which is given
in the previous section, $u_\Lambda$ and $v_{\bar{\Lambda}}$ are the
spinors for the $\Lambda$ and $\bar{\Lambda}$ baryons, respectively.
The $\Gamma_{1,2}$ is vertex as $1$, $\gamma_5$, or $\epsilon\!\!\!/$
for scalar, pseudoscalar and vector mesons, respectively. Here, we need an additional coupling constant $g_{NK\Lambda}=13.926$ for $0^-(1^{--})$ state~\cite{Dong:2017rmg}. The $q$ and
$m_\Lambda$ are the momentum and mass of the exchanged $\Lambda$
baryon.  $f_i(q^2)$ is the form factor as introduced in the previous
section. The $\zeta$ is a sign from the difference between baryon and
antibaryon as G-parity rule. It does not affect the result because the
above interaction always appears in a pair. 

Obviously, the above treatment is not enough to include all
annihilation effect. A phenomenological treatment is often introduced
in the
literature~\cite{Mull:1994gz,Bryan:1968ns,Dover:1980pd,Kohno:1986fk,Richard:1982zr,Maruyama:1985zm,Bydzovsky:1991mc,Haidenbauer:1991kt}.
In the current work, we introduce an additional imaginary optical
potential into $V_{el}$ with following parameterization in coordinate
space as in Ref.~\cite{Mull:1994gz}
\begin{eqnarray}
V_{opt}=iWe^{-\frac{r^2}{2r_0^2}}.
\end{eqnarray}
To insert such optical potential into our qBSE approach, we need to transform it into momentum space by the Fourier transformation as,
\begin{eqnarray}
  {V}_{opt}(q^2)=4\pi\sqrt{\frac{\pi}{2}} iW r_0^3e^{q^2 r_0^2/2}
  \bar{u}_\Lambda u_\Lambda \bar{u}_{\bar{\Lambda}} u_{\bar{\Lambda}},
\end{eqnarray}
where the $q$ is the four momentum as for the exchanged mesons.  Due to lack of the experimental data for the
$\Lambda\bar{\Lambda}$ interaction, the parameters was not so well
determined as the $N\bar{N}$ interaction. In
Ref~\cite{Haidenbauer:1991kt}, the $p\bar{p}\to\Lambda\bar{\Lambda}$
process was studied, and the parameters are determined as $W\approx-1$
~GeV and $r_0\approx0.3$~fm, which will be adopted in the current
calculation.   
Such values  are similar to those in the
nucleon-nucleon interaction as adopted in Ref.~\cite{Mull:1994gz}, 
$W=-1$~GeV and $r_0=0.4$~fm,  which are 
also close to the values adopted in Ref.~\cite{Bydzovsky:1991mc}.

The positions of the poles of the bound states with different
$\alpha$ are listed in
Table~\ref{tab1}. The real part of the
$M_{th}-z$ presented in the table is the binding energy of the
$\Lambda\bar{\Lambda}$ molecular states. After including the annihilation effect, the
poles appear at $\alpha$ around 1.9, 0.8 and 0.4  with different form
factors, respectively. Compared with the results in Fig.~\ref{total},
larger  $\alpha$  is required to produce  molecular states from the interaction.
It suggests that the attraction of the interaction becomes weaker, which
needs a  larger $\alpha$ to compensate. However, the changes of the mass gap
between $0^-$ and $1^-$ states is slight. For most cases, the mass of the $1^-$ state
even becomes further larger than that of the $0^-$ state, with mass gaps about
11, 8, and 5 MeV for three form factors, which is more consistent with the
assignment of two states as $X(2235)$ and $\eta(2225)$. In
Table~\ref{tab1}, we also present the results without the coupled-channel
effect, that is, only with the imaginary optical potential.  The result
suggests that coupled-channel effect on the mass is obvious. The mass
decreases by about 5  and 10 MeV  for $0^-$ and $1^-$ states after
including the coupled-channel effect.

 \renewcommand\tabcolsep{0.2cm}
 \renewcommand{\arraystretch}{1.3}
\begin{table}[h!]
\begin{center}
\caption{
The position of the poles of $0^-$ and $1^-$  states with different $\alpha$. 
The $M_{th}-z$ means mass of the
$\Lambda\bar{\Lambda}$ threshold $M_{th}$ subtracted by the position of the
pole $z$, in a unit of MeV. The $f_i$ means the results
with different types of form factors. The first and second lines for every $\alpha$ is
for the results without and with couple-channel effect. For $1^-$ state, the results  with couple-channel effect  except the $K\bar{K}$ channel are listed in third line.
\label{tab1}}
	\begin{tabular}{c|crcrcr}\bottomrule[1.5pt]
	&\multicolumn{2}{c}{$f_1$}&\multicolumn{2}{c}{$f_2$}&\multicolumn{2}{c}{$f_3$}\\\hline
$J^P$   &  $\alpha$  &  $M_{th}-z$$\ \ \ \ $  &  $\alpha$  & $M_{th}-z$$\ \ \ \ $&  $\alpha$  &  $M_{th}-z$$\ \ \ \ $ \\\hline
$0^-$
& 1.9       & $11.9+84i$   &   0.8     & $8.42+38i$    &  0.4   & $6.21+25i$   \\
&   & $6.35+84i$   &   & $3.41+37i$    &  & $3.17+22i$  \\
& 2.0       & $22.8+89i$   &   0.9  & $16.8+43i$    & 0.5     & $13.2+30i$    \\
&   & $17.2+91i$   &  & $10.3+41i$    & & $9.58+27i$   \\
&  2.1      & $34.4+94i$   &   1.0       & $25.9+48i$    &   0.6    & $21.7+34i$   \\
&   & $28.6+96i$   & & $18.6+46i$    &    & $16.2+31i$  \\
\hline
$1^-$
&  2.0      &$14.5+85i$       &    0.9    & $14.6+41i$  &    0.5 & $12.2+27i$  \\
&   &$--\quad$       &   & $--\quad$  &    & $--\quad$     \\
&   &$6.67+97i$       &   & $2.41+46i$  &    & $4.26+29i$     \\
&  2.1     & $25.5+\ \ 90i$      &  1.0& $22.2+44i$  &    0.6 & $19.3+32i$   \\
&      & $3.6+114i$     &    & $1.21+55i$  &     & $0.67+38i$   \\
&    & $17.4+104i$     &   & $9.02+50i$  & & $9.62+35i$   \\
&2.2  & $37.5+\ \ 95i$  &  1.1  & $37.7+48i$  &   0.7    & $27.6+36i$    \\
&       & $8.3+132i$      &       & $12.5+65i$  &        & $3.1+47i$    \\
&  & $22.8+113i$     &  & $26.4+58i$  & & $15.3+41i$     \\
\toprule[1.5pt]
\end{tabular}
\end{center}
\end{table}

Another obvious variation after including the annihilation effect is that
the poles leave the real axis and the states acquire widths. The imaginary
part of $M_{th}-z$ corresponds to the half of the width of the states. The
current result suggests  large widths for both $0^-$ and $1^-$ states,
about 200, 100, and 60 MeV for three form factors, respectively. It is
consistent with the experimental observation of the $X(2239)$ and
$\eta(2225)$ with widths of $139.8\pm12.3\pm20.6$
MeV~\cite{Ablikim:2018iyx} and $185^{+40}_{-20}$
MeV~\cite{Tanabashi:2018oca}, respectively. Here, we also consider the
results without the coupled-channel effect. One can find that the
variations of the widths for two states and different form factors are from several to about ten MeV.  Considering the widths are several dozens of MeV, the variations of widths from the coupled-channel effect considered here are relatively small. The widths are mainly from the imaginary potential.  For the $1^-$ state, we present the results with and without $K\bar{K}$ channel, the result suggests that $K\bar{K}$ channel provides a width comparable with all the contribution from other channels.

\section{Summary and discussion}\label{Sec: summary}

The molecular state composed of a baryon and an antibaryon is an
interesting topic in the study of  exotic mesons. In the present work, we
study the possibility to assign the newly observed $X(2239)$ as a
$\Lambda\bar{\Lambda}$ molecular state in the qBSE approach.  The potential
kernel of the $\Lambda\bar{\Lambda}$ interaction is constructed within the
one-boson-exchange model with the help of the Feynman rule,  and the annihilation effect was introduced through introducing the coupled-channel effect and optical potential.  After decomposition on spin parity, the bound state
can be found by studying the pole of  the scattering amplitude.

Two bound states with spin parities $J^P=0^-$ and $1^-$ are produced from
the $\Lambda\bar{\Lambda}$ interaction.  Our results suggest that these two
bound states are both close to the $\Lambda\bar{\Lambda}$ threshold.
Before the observation of the $X(2239)$, there existed only one possible
state $Y(2175)$ near the $\Lambda\bar{\Lambda}$ interaction,  so it is
often assigned as the $1^-$ molecular state. However,  a binding energy
larger than 50 MeV is required for this assignment.  Now, the $X(2239)$ was
observed in the  $K^+K^-$ channel, and is just on the
$\Lambda\bar{\Lambda}$ threshold if the experimental uncertainty is
considered.  Besides,  in Ref.~\cite{Dong:2017rmg},   the study of  strong
decays of the $\Lambda\bar{\Lambda}$ bound state was performed, and it was
found that the dominant decay channel of the $1^-$ state is the $K\bar{K}$
channel, which is just the observation  channel of the $X(2239)$ at BESIII. In the current work, the $K\bar{K}$ channel is also found to provide considered width in all channels considered.
Hence,  it is more suitable to assign these two states with spin parities
$1^-$ and $0^-$ from the $\Lambda\bar{\Lambda}$ interaction to the
$X(2239)$ and the $\eta(2225)$, respectively.

 We  discuss the effect of each exchange on producing the bound state.
 Among the five exchanges including $\eta$, $\eta'$, $\phi$, $\omega$, and
 $\sigma$ exchanges, the $\omega$ and $\sigma$ exchanges, especially the
 former, play the  most important role to produce two bound states. Such
 conclusion is consistent with the previous studies in
 Refs.~\cite{Zhao:2013ffn,Lee:2011rka}. We also check the effect of
 different choices of form factors on the conclusion. The behaviors of the
 results  with three types of form factors are analogous to each other.  If
 we recall that the cutoff is a free parameter, it suggests that the same
 conclusion can be reached with different choices of the form factors.

In the $N\bar{N}$ interaction, the annihilation is an important
topic~\cite{Klempt:2005pp,Richard:2019dic,Amsler:1997up}. In the current
work, we include the annihilation effect by following the procedure in
Ref.~\cite{Mull:1994gz}. The coupled-channel effect from the  two-meson
intermediation and an imaginary optical potential are introduced and insert
to our qBSE approach. However, unlike the $N\bar{N}$ interaction, the
experimental information about the $\Lambda\bar{\Lambda}$ interaction is
scarce. Hence, we choose the parameters from fitting the data of the
$p\bar{p}\to\Lambda\bar{\Lambda}$ process~\cite{Haidenbauer:1991kt}.  The
calculation suggests that the variations of the poles are very large. The
cutoff should be increased to give the molecular states. And a large width
about 100 MeV is also produced from the annihilation especially the optical
potential.  However,  our conclusion from the single-channel calculation with one-boson exchange is
unchanged qualitatively after the annihilation effect is considered.  More
exact determination of such effect requires more experimental data and
theoretical analysis.

In the current work, we propose that the observation of the $X(2239)$ at
BESIII provides a more suitable candidate of the $\Lambda\bar{\Lambda}$
molecular state with spin parity $1^-$.   In the charmed sector, the
$\Lambda_c\bar{\Lambda}_c$ molecular state  was also studied in  theory,
and was assigned as the $Y(4630)$ by some
authors~\cite{Lee:2011rka}. As in the hidden-strange sector,
many states were observed near the $\Lambda_c\bar{\Lambda}_c$ threshold,
including $Y(4630)$, $Y(4660)$ and a structure at 4625 MeV observed at
Belle very recently~\cite{Jia:2019gfe}, which attracts many attentions of
theorists~\cite{Guo:2010tk,Cao:2019wwt}.  More comprehensive investigation about the
$\Lambda_{(c)}\bar{\Lambda}_{(c)}$ interaction in both theory and
experiment is important to understand  these structures. 

The
conclusion of the current work is based on the assumption that the
$X(2239)$  is below the $\Lambda\bar{\Lambda}$ threshold. Though the
experimental mass with uncertainties can reach the region below the
threshold, the nominal value is above the threshold. If it is still true
with smaller uncertainties, the $X(2239)$  can not be explained as a
molecular state from the $\Lambda\bar{\Lambda}$ interaction as suggested in
the current work. Besides, the current work is based on the experimental results about the  $X(2239)$ released by the BESIII Collaboration~\cite{Ablikim:2018iyx}. More experimental data and careful analysis are  necessary to confirm whether the
$X(2239)$ and $Y(2175)$ are different states.  In Ref.~\cite{Chen:2020xho}, a fitting of the experimental data points for process $e^+e^-\to K^+K^-$ at BESIII suggests that the structure can be reproduced with the interference between a states near 2200 MeV, which is also much larger than usual mass of $Y(2175)$, and the background without a real state near 2.24 GeV.   However, the current experimental data are not enough to give a confirmative  conclusion.     Hence, a more precise
measurement of the mass of the $X(2239)$ is very important to confirm such
assignment~\cite{Wang:2019uwk}.

\acknowledgments

This project is partially supported by the National Natural Science
Foundation of China (Grants No.11675228, No. 11775050, and No. 61871124),
the Fundamental Research Funds for the Central Universities, the national
defense Pre-Research foundation of China, by State Key Laboratory of
Acoustics, Chinese Academy of Sciences (No: SKLA201604), by Science and
Technology on Sonar Laboratory, by the Stable Supporting Fund of Acoustic
Science and Technology Laboratory.

\end{document}